\pdfoutput=1

\documentclass[12pt,a4paper]{article}

\usepackage{ifthen} 
\newboolean{pdflatex}
\setboolean{pdflatex}{true} 

\newboolean{articletitles}
\setboolean{articletitles}{true} 

\newboolean{uprightparticles}
\setboolean{uprightparticles}{false} 

\usepackage{microtype}
\usepackage{lineno}  
\usepackage{xspace} 

\usepackage{graphicx}  
\usepackage{color}
\usepackage{colortbl}
\graphicspath{{./figs/}} 

\usepackage{amsmath} 
\usepackage{amssymb}
\usepackage{amsfonts}
\usepackage{upgreek} 
\usepackage{calc}

\newcommand*\patchAmsMathEnvironmentForLineno[1]{%
\expandafter\let\csname old#1\expandafter\endcsname\csname #1\endcsname
\expandafter\let\csname oldend#1\expandafter\endcsname\csname
end#1\endcsname
 \renewenvironment{#1}%
   {\linenomath\csname old#1\endcsname}%
   {\csname oldend#1\endcsname\endlinenomath}%
}
\newcommand*\patchBothAmsMathEnvironmentsForLineno[1]{%
  \patchAmsMathEnvironmentForLineno{#1}%
  \patchAmsMathEnvironmentForLineno{#1*}%
}
\AtBeginDocument{%
\patchBothAmsMathEnvironmentsForLineno{equation}%
\patchBothAmsMathEnvironmentsForLineno{align}%
\patchBothAmsMathEnvironmentsForLineno{flalign}%
\patchBothAmsMathEnvironmentsForLineno{alignat}%
\patchBothAmsMathEnvironmentsForLineno{gather}%
\patchBothAmsMathEnvironmentsForLineno{multline}%
}

\usepackage{hyperref}    
\usepackage[all]{hypcap} 

\newcommand{\comment}[1]{}

\def\lhcb {\mbox{LHCb}\xspace}
\def\ux85 {\mbox{UX85}\xspace}

\ifthenelse{\boolean{uprightparticles}}%
{

 \def\Pmu         {\ensuremath{\upmu}\xspace}

 \def\Ppi         {\ensuremath{\uppi}\xspace}

 \def\Ppsi        {\ensuremath{\uppsi}\xspace}

 \def\PDelta      {\ensuremath{\Delta}\xspace}                 
 \def\PXi      {\ensuremath{\Xi}\xspace}                 
 \def\PLambda      {\ensuremath{\Lambda}\xspace}                 
 \def\PSigma      {\ensuremath{\Sigma}\xspace}                 
 \def\POmega      {\ensuremath{\Omega}\xspace}                 
 \def\PUpsilon      {\ensuremath{\Upsilon}\xspace}                 
 
 \def\PB      {\ensuremath{\mathrm{B}}\xspace}                 
                  
 \def\PD      {\ensuremath{\mathrm{D}}\xspace}

 \def\PJ      {\ensuremath{\mathrm{J}}\xspace}                 
 \def\PK      {\ensuremath{\mathrm{K}}\xspace}

 \def\Pb      {\ensuremath{\mathrm{b}}\xspace}                 
 \def\Pc      {\ensuremath{\mathrm{c}}\xspace}

 \def\Pi      {\ensuremath{\mathrm{i}}\xspace}

 \def\Ps      {\ensuremath{\mathrm{s}}\xspace}

}
{

 \def\Pmu         {\ensuremath{\mu}\xspace}

 \def\Ppi         {\ensuremath{\pi}\xspace}

 \def\Ppsi        {\ensuremath{\psi}\xspace}                 
                  
 \mathchardef\PDelta="7101
 \mathchardef\PXi="7104
 \mathchardef\PLambda="7103
 \mathchardef\PSigma="7106
 \mathchardef\POmega="710A
 \mathchardef\PUpsilon="7107
                  
 \def\PB      {\ensuremath{B}\xspace}                 
                  
 \def\PD      {\ensuremath{D}\xspace}

 \def\PJ      {\ensuremath{J}\xspace}                 
 \def\PK      {\ensuremath{K}\xspace}

 \def\Pb      {\ensuremath{b}\xspace}                 
 \def\Pc      {\ensuremath{c}\xspace}

 \def\Pi      {\ensuremath{i}\xspace}

 \def\Ps      {\ensuremath{s}\xspace}

}

\def\mup        {\ensuremath{\Pmu^+}\xspace}
\def\mun        {\ensuremath{\Pmu^-}\xspace} 

\def\squark    {\ensuremath{\Ps}\xspace}

\def\cquark    {\ensuremath{\Pc}\xspace}

\def\bquark    {\ensuremath{\Pb}\xspace}

\def\pion  {\ensuremath{\Ppi}\xspace}

\def\pip   {\ensuremath{\pion^+}\xspace}
\def\pim   {\ensuremath{\pion^-}\xspace}

\def\kaon  {\ensuremath{\PK}\xspace}
  \def\Kbar  {\kern 0.2em\overline{\kern -0.2em \PK}{}\xspace}

\def\Kz    {\ensuremath{\kaon^0}\xspace}
\def\Kzb   {\ensuremath{\Kbar^0}\xspace}
\def\KzKzb {\ensuremath{\Kz \kern -0.16em \Kzb}\xspace}
\def\Kp    {\ensuremath{\kaon^+}\xspace}
\def\Km    {\ensuremath{\kaon^-}\xspace}

\def\KpKm  {\ensuremath{\Kp \kern -0.16em \Km}\xspace}

\def\Kstarz  {\ensuremath{\kaon^{*0}}\xspace}

  \def\Dbar    {\kern 0.2em\overline{\kern -0.2em \PD}{}\xspace}
\def\D       {\ensuremath{\PD}\xspace}

\def\Dz      {\ensuremath{\D^0}\xspace}
\def\Dzb     {\ensuremath{\Dbar^0}\xspace}
\def\DzDzb   {\ensuremath{\Dz {\kern -0.16em \Dzb}}\xspace}
\def\Dp      {\ensuremath{\D^+}\xspace}
\def\Dm      {\ensuremath{\D^-}\xspace}

\def\DpDm    {\ensuremath{\Dp {\kern -0.16em \Dm}}\xspace}

\def\Dsm     {\ensuremath{\D^-_\squark}\xspace}

\def\B       {\ensuremath{\PB}\xspace}
\def\Bbar    {\ensuremath{\kern 0.18em\overline{\kern -0.18em \PB}{}}\xspace}

\def\Bd      {\ensuremath{\B^0}\xspace}
\def\Bs      {\ensuremath{\B^0_\squark}\xspace}
\def\Bsb     {\ensuremath{\Bbar^0_\squark}\xspace}

\def\jpsi     {\ensuremath{{\PJ\mskip -3mu/\mskip -2mu\Ppsi\mskip 2mu}}\xspace}
\def\psitwos  {\ensuremath{\Ppsi{(2S)}}\xspace}

  \def\Y#1S{\ensuremath{\PUpsilon{(#1S)}}\xspace}

\def\L {\ensuremath{\PLambda}\xspace}
\def\Lbar {\ensuremath{\kern 0.1em\overline{\kern -0.1em\PLambda}}\xspace}

\def\Lb      {\ensuremath{\L^0_\bquark}\xspace}

\newcommand{\decay}[2]{\ensuremath{#1\!\to #2}\xspace}         

\def\to                 {\ensuremath{\rightarrow}\xspace}

\def\CP                {\ensuremath{C\!P}\xspace}

\def\BsToPhimm    {\decay{\Bs}{\phi\mup\mun}}

\def\BdToKstmm    {\decay{\Bd}{\Kstarz\mup\mun}}

\def\BsToJPsiPhi  {\decay{\Bs}{\jpsi\phi}}
\def\BdToJPsiKst  {\decay{\Bd}{\jpsi\Kstarz}}

\def\AT#1     {\ensuremath{A_{\mathrm{T}}^{#1}}\xspace}           

\def\ctl       {\ensuremath{\cos{\theta_\ell}}\xspace}
\def\ctk       {\ensuremath{\cos{\theta_K}}\xspace}
\def\thetal       {\ensuremath{\theta_\ell}\xspace}
\def\thetak       {\ensuremath{\theta_K}\xspace}

\def\C#1      {\ensuremath{\mathcal{C}_{#1}}\xspace}                       
\def\Cp#1     {\ensuremath{\mathcal{C}_{#1}^{'}}\xspace}                    
\def\Ceff#1   {\ensuremath{\mathcal{C}_{#1}^{\mathrm{(eff)}}}\xspace}        
\def\Cpeff#1  {\ensuremath{\mathcal{C}_{#1}^{'\mathrm{(eff)}}}\xspace}       
\def\Ope#1    {\ensuremath{\mathcal{O}_{#1}}\xspace}                       
\def\Opep#1   {\ensuremath{\mathcal{O}_{#1}^{'}}\xspace}                    

\newcommand{\tev}{\ensuremath{\mathrm{\,Te\kern -0.1em V}}\xspace}
\newcommand{\gev}{\ensuremath{\mathrm{\,Ge\kern -0.1em V}}\xspace}
\newcommand{\mev}{\ensuremath{\mathrm{\,Me\kern -0.1em V}}\xspace}
\newcommand{\kev}{\ensuremath{\mathrm{\,ke\kern -0.1em V}}\xspace}
\newcommand{\ev}{\ensuremath{\mathrm{\,e\kern -0.1em V}}\xspace}
\newcommand{\gevc}{\ensuremath{{\mathrm{\,Ge\kern -0.1em V\!/}c}}\xspace}
\newcommand{\mevc}{\ensuremath{{\mathrm{\,Me\kern -0.1em V\!/}c}}\xspace}
\newcommand{\gevcc}{\ensuremath{{\mathrm{\,Ge\kern -0.1em V\!/}c^2}}\xspace}
\newcommand{\gevgevcccc}{\ensuremath{{\mathrm{\,Ge\kern -0.1em V^2\!/}c^4}}\xspace}
\newcommand{\mevcc}{\ensuremath{{\mathrm{\,Me\kern -0.1em V\!/}c^2}}\xspace}

\def\mum  {\ensuremath{\,\upmu\rm m}\xspace}

\def\invfb   {\ensuremath{\mbox{\,fb}^{-1}}\xspace}

\newcommand{\chisq}{\ensuremath{\chi^2}\xspace}

\def\gsim{{~\raise.15em\hbox{$>$}\kern-.85em
          \lower.35em\hbox{$\sim$}~}\xspace}
\def\lsim{{~\raise.15em\hbox{$<$}\kern-.85em
          \lower.35em\hbox{$\sim$}~}\xspace}

\def\pt         {\mbox{$p_{\rm T}$}\xspace}
\def\et         {\mbox{$E_{\rm T}$}\xspace}

\def\evtgen     {\mbox{\textsc{EvtGen}}\xspace}
\def\pythia     {\mbox{\textsc{Pythia}}\xspace}

\def\geant      {\mbox{\textsc{Geant4}}\xspace}

\def\photos     {\mbox{\textsc{Photos}}\xspace}

\def\tell1  {TELL1\xspace}
\def\ukl1   {UKL1\xspace}

\newcommand{\eg}{\mbox{\itshape e.g.}\xspace}
\newcommand{\ie}{\mbox{\itshape i.e.}}

\usepackage{cite} 
\usepackage{mciteplus}
\usepackage{multirow}

\usepackage{longtable} 

\begin{document}

\renewcommand{\thefootnote}{\fnsymbol{footnote}}
\setcounter{footnote}{1}


\begin{titlepage}
\pagenumbering{roman}

\vspace*{-1.5cm}
\centerline{\large EUROPEAN ORGANIZATION FOR NUCLEAR RESEARCH (CERN)}
\vspace*{1.5cm}
\hspace*{-0.5cm}
\begin{tabular*}{\linewidth}{lc@{\extracolsep{\fill}}r}
\ifthenelse{\boolean{pdflatex}}
{\vspace*{-2.7cm}\mbox{\!\!\!\includegraphics[width=.14\textwidth]{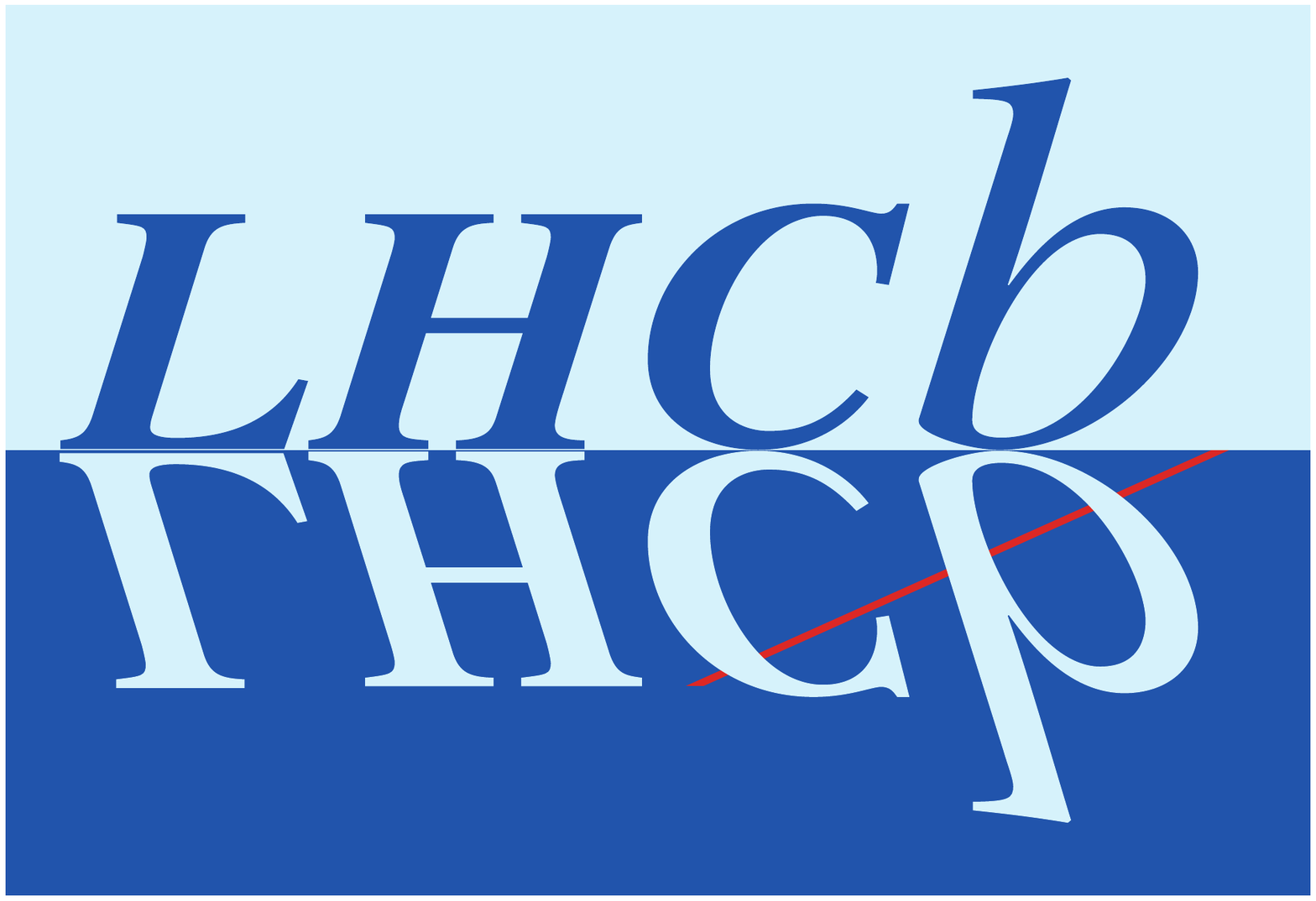}} & &}%
{\vspace*{-1.2cm}\mbox{\!\!\!\includegraphics[width=.12\textwidth]{lhcb-logo.eps}} & &}%
\\
 & & CERN-PH-EP-2013-078 \\  
 & & LHCb-PAPER-2013-017 \\  
 & & June 19, 2013 \\ 
 & & \\
\end{tabular*}

\vspace*{2.0cm}

{\bf\boldmath\huge
\begin{center}
Differential branching fraction\\
and angular analysis of\\ the decay $\BsToPhimm$
\end{center}
}

\vspace*{1.0cm}

\begin{center}
The LHCb collaboration\footnote{Authors are listed on the following pages.}
\end{center}

\vspace{\fill}

\begin{abstract}
  \noindent
The determination of the differential branching fraction and 
the first angular analysis of the decay $\BsToPhimm$ are presented using data, corresponding to an integrated luminosity of $1.0\invfb$, collected by the LHCb experiment at $\sqrt{s}=7\tev$. 
The differential branching fraction is determined in bins of $q^2$, the invariant dimuon mass squared. 
Integration over the full $q^2$ range yields a total branching fraction of 
\mbox{${\cal B}(\Bs\rightarrow\phi\mu^+\mu^-) = \left(7.07\,^{+0.64}_{-0.59}\pm 0.17 \pm 0.71\right)\times 10^{-7}$}, 
where the first uncertainty is statistical, the second systematic, and the third originates from the branching fraction of the normalisation channel. 
An angular analysis is performed to determine the angular observables $F_{\rm L}$, $S_3$, $A_6$, and $A_9$. 
The observables are consistent with Standard Model expectations. 
\end{abstract}

\vspace*{1.0cm}

\begin{center}
  Submitted to JHEP
\end{center}

\vspace{\fill}

{\footnotesize 
\centerline{\copyright~CERN on behalf of the \lhcb collaboration, license \href{http://creativecommons.org/licenses/by/3.0/}{CC-BY-3.0}.}}
\vspace*{2mm}

\end{titlepage}


\newpage
\setcounter{page}{2}
\mbox{~}
\newpage

\centerline{\large\bf LHCb collaboration}
\begin{flushleft}
\small
R.~Aaij$^{40}$, 
C.~Abellan~Beteta$^{35,n}$, 
B.~Adeva$^{36}$, 
M.~Adinolfi$^{45}$, 
C.~Adrover$^{6}$, 
A.~Affolder$^{51}$, 
Z.~Ajaltouni$^{5}$, 
J.~Albrecht$^{9}$, 
F.~Alessio$^{37}$, 
M.~Alexander$^{50}$
% ...

{\footnotesize \it
% ...
$ ^{5}$Clermont Universit\'{e}, Universit\'{e} Blaise Pascal, CNRS/IN2P3, LPC, Clermont-Ferrand, France\\
$ ^{6}$CPPM, Aix-Marseille Universit\'{e}, CNRS/IN2P3, Marseille, France\\
% ...
$ ^{9}$Fakult\"{a}t Physik, Technische Universit\"{a}t Dortmund, Dortmund, Germany\\
% ...
$ ^{35}$Universitat de Barcelona, Barcelona, Spain\\
$ ^{36}$Universidad de Santiago de Compostela, Santiago de Compostela, Spain\\
$ ^{37}$European Organization for Nuclear Research (CERN), Geneva, Switzerland\\
% ...
$ ^{40}$Nikhef National Institute for Subatomic Physics, Amsterdam, The Netherlands\\
% ...
$ ^{45}$H.H. Wills Physics Laboratory, University of Bristol, Bristol, United Kingdom\\
% ...
$ ^{50}$School of Physics and Astronomy, University of Glasgow, Glasgow, United Kingdom\\
$ ^{51}$Oliver Lodge Laboratory, University of Liverpool, Liverpool, United Kingdom\\
% ...
$ ^{n}$LIFAELS, La Salle, Universitat Ramon Llull, Barcelona, Spain\\
% ...
$ ^{s}$Scuola Normale Superiore, Pisa, Italy\\
}
\end{flushleft}

\cleardoublepage


\renewcommand{\thefootnote}{\arabic{footnote}}
\setcounter{footnote}{0}



\pagestyle{plain} 
\setcounter{page}{1}
\pagenumbering{arabic}


%

\section{Introduction}
\label{sec:Introduction}
The $\BsToPhimm$ ($\decay{\phi}{\Kp\Km}$) decay\footnote{The inclusion of charge conjugated processes is implied throughout this paper.} involves a $b\rightarrow s$ quark transition and therefore constitutes a flavour changing neutral current (FCNC) process. 
Since FCNC processes are forbidden at tree level in the Standard Model (SM), the decay 
is mediated by higher order (box and penguin) diagrams. 
In scenarios beyond the SM new particles can 
affect both the branching fraction of the decay and the angular distributions of the decay products. 

The angular configuration of the $\Kp\Km\mup\mun$ system is defined by the decay angles $\thetak$, $\thetal$, and $\Phi$. 
Here, $\thetak$ ($\thetal$) denotes the angle of the $\Km$ ($\mun$) with respect to the direction of flight of the \Bs\ meson in the $\Kp\Km$ ($\mup\mun$)  
centre-of-mass frame, 
and $\Phi$ denotes the relative angle of the 
$\mup\mun$ and the $\Kp\Km$ decay planes 
in the $\Bs$ meson centre-of-mass frame~\cite{bib:hiller}. 
In contrast to the decay $\BdToKstmm$, the final state of the decay $\BsToPhimm$ is not flavour specific. 
The differential decay rate, 
depending on the decay angles and the invariant mass squared of the dimuon system is given by
\begin{align}
\frac{1}{{\rm d}\Gamma/{\rm d}q^2}\frac{{\rm d}^4\Gamma}{{\rm d}q^2{\rm d}\ctl {\rm d}\ctk {\rm d}\Phi}
=
\frac{9}{32\pi} 
& \bigl[ 
S_1^s\sin^2\theta_K+S_1^c\cos^2\theta_K\nonumber\\[-7pt]
 & +S_2^s\sin^2\theta_K\cos 2\theta_\ell + S_2^c\cos^2\theta_K\cos 2\theta_\ell\nonumber\\
 & + {S_3} \sin^2\theta_K\sin^2\theta_\ell \cos 2\Phi
+ {S_4} \sin 2\theta_K\sin 2\theta_\ell \cos \Phi\nonumber\\
 & + {A_5} \sin 2\theta_K\sin \theta_\ell \cos \Phi
+  {A_6} \sin^2\theta_K\cos\theta_\ell\nonumber\\
 & + {S_7} \sin 2\theta_K \sin\theta_\ell \sin\Phi
+  {A_8} \sin 2\theta_K \sin 2\theta_\ell \sin\Phi\nonumber\\
 & + {A_9} \sin^2\theta_K \sin^2\theta_\ell \sin 2\Phi
\bigr],\label{eq:untagged}
\end{align}
where equal numbers of produced \Bs\ and \Bsb\ mesons are assumed~\cite{bib:altmannshofer}. 
The $q^2$-dependent angular observables $S_i^{(s,c)}$ and $A_i$ correspond to \CP\ averages and \CP\ asymmetries, respectively. 
Integrating Eq.~\ref{eq:untagged} over two angles, under the assumption of massless leptons, results in three distributions, each depending on one decay angle
\begin{align}
\frac{1}{{\rm d}\Gamma/{\rm d}q^2}\frac{{\rm d}^2\Gamma }{{\rm d}q^2\,{\rm d}\cos\theta_K} 
&= \frac{3}{4}(1-{F_{\rm L}}) (1 - \cos^2\theta_K) + \frac{3}{2} {F_{\rm L}} \cos^2\theta_K,\label{eq:ctk}\\
\frac{1}{{\rm d}\Gamma/{\rm d}q^2}\frac{{\rm d}^2\Gamma }{{\rm d}q^2\,{\rm d}\cos\theta_\ell} 
&= \frac{3}{8}(1-{F_{\rm L}}) ( 1+\cos^2\theta_\ell ) + \frac{3}{4} {F_{\rm L}} ( 1-\cos^2\theta_\ell ) + \frac{3}{4} {A_6}\cos\theta_\ell,\label{eq:ctl}\\
\frac{1}{{\rm d}\Gamma/{\rm d}q^2}\frac{{\rm d}^2\Gamma }{{\rm d}q^2\,{\rm d}\Phi} 
&= \frac{1}{2\pi} + \frac{1}{2\pi}{S_3}\cos2\Phi + \frac{1}{2\pi} {A_9} \sin 2\Phi\label{eq:phi},
\end{align}
which retain sensitivity to the angular observables $F_{\rm L}(=S_1^c=-S_2^c)$, $S_3$, $A_6$, and $A_9$. 
Of particular interest is the $T$-odd asymmetry $A_9$ where possible large \CP -violating phases from contributions beyond the SM would not be suppressed by small strong phases~\cite{bib:hiller}. 

This paper presents a measurement of the differential branching fraction and the angular observables $F_{\rm L}$, $S_3$, $A_6$, and $A_9$ in six bins of $q^2$. 
In addition, the total branching fraction is determined. 
The data used in the analysis were recorded by the LHCb experiment in 2011 in $pp$ collisions at $\sqrt{s}=7\tev$ and correspond to an integrated luminosity of $1.0\invfb$.

\section{The LHCb detector}
\label{sec:detector}
The \lhcb detector~\cite{Alves:2008zz} is a single-arm forward
spectrometer covering the \mbox{pseudorapidity} range $2<\eta <5$,
designed for the study of particles containing \bquark or \cquark
quarks. The detector includes a high precision tracking system
consisting of a silicon-strip vertex detector surrounding the $pp$
interaction region, a large-area silicon-strip detector located
upstream of a dipole magnet with a bending power of about
$4{\rm\,Tm}$, and three stations of silicon-strip detectors and straw
drift tubes placed downstream. 
The combined tracking system provides a momentum measurement with
relative uncertainty that varies from 0.4\% at 5\gevc to 0.6\% at 100\gevc,
and impact parameter (IP) resolution of 20\mum for
tracks with high transverse momentum. 
Charged hadrons are identified
using two ring-imaging Cherenkov detectors. Photon, electron and
hadron candidates are identified by a calorimeter system consisting of
scintillating-pad and preshower detectors, an electromagnetic
calorimeter and a hadronic calorimeter. Muons are identified by a
system composed of alternating layers of iron and multiwire
proportional chambers. The LHCb trigger system~\cite{Aaij:2012me} consists of a
hardware stage, based on information from the calorimeter and muon
systems, followed by a software stage which applies a full event
reconstruction.

Simulated signal event samples are generated to determine the trigger, reconstruction and selection efficiencies. 
Exclusive samples are analysed to estimate possible backgrounds. 
The simulation generates $pp$ collisions using 
\pythia~6.4~\cite{Sjostrand:2006za} with a specific \lhcb
configuration~\cite{LHCb-PROC-2010-056}.  Decays of hadronic particles
are described by \evtgen~\cite{Lange:2001uf} in which final state
radiation is generated using \photos~\cite{Golonka:2005pn}. The
interaction of the generated particles with the detector and its
response are implemented using the \geant
toolkit~\cite{Allison:2006ve, *Agostinelli:2002hh} as described in
Ref.~\cite{LHCb-PROC-2011-006}. 
Data driven corrections are applied to the simulated events to account for differences between data and simulation. 
These include the IP resolution, 
tracking efficiency, and particle identification performance. 
In addition, 
simulated events are reweighted depending on 
the transverse momentum ($\pt$) of the \Bs\ meson, the vertex fit quality, and the track multiplicity 
to match distributions of control samples from data.

\section{Selection of signal candidates}
\label{sec:selection}
Signal candidates are accepted if they are triggered by particles of the $\BsToPhimm$ ($\decay{\phi}{\Kp\Km}$) final state. 
The hardware trigger requires 
either a high transverse momentum muon or muon pair, or a high transverse energy ($\et$) hadron. 
The first stage of the software trigger selects events containing a muon (or hadron) with $\pt >0.8\gevc$ ($\et >1.5\gevc$) and a minimum IP 
with respect to all primary interaction vertices in the event 
of $80\mum$ ($125\mum$). 
In the second stage of the software trigger 
the tracks of two or more final state 
particles are required to form a vertex that is significantly 
displaced from all primary vertices~(PVs)   
in the event. 

Candidates are selected 
if they pass a loose preselection that requires the kaon and muon tracks to have a large $\chi^2_{\rm IP}$ ($>9$) with respect to the PV. 
The $\chi^2_{\rm IP}$ is defined as the difference between the $\chi^2$ of the PV reconstructed with and without the considered particle. 
The four tracks forming a \Bs\ candidate are fit to a common vertex, which is required to be of good quality ($\chi^2_{\rm vtx}<30$) and well separated from the PV ($\chi^2_{\rm FD}>121$, 
where FD denotes the flight distance). 
The angle between the \Bs\ momentum vector and the vector connecting the PV with the \Bs\ decay vertex is required to be small. 
Furthermore, \Bs\ candidates are required to have a small IP 
with respect to the PV ($\chi^2_{\rm IP}<16$).  
The invariant mass of the $\Kp\Km$ system is required to be within $12\mevcc$ of the known $\phi$ mass~\cite{PDG2012}. 

To further reject combinatorial background events, a boosted decision tree~(BDT)~\cite{Breiman} 
using the AdaBoost algorithm~\cite{AdaBoost} is applied. 
The BDT training uses $\BsToJPsiPhi$ $(\jpsi\to\mup\mun)$ candidates as proxy for the signal, and candidates in the $\Bs\to\phi\mup\mun$ mass sidebands 
($5100 <m(\Kp\Km\mup\mun)< 5166\mevcc$ and $5566 <m(\Kp\Km\mup\mun)< 5800\mevcc$)
as background. 
The input variables of the BDT are 
the $\chi^2_{\rm IP}$ of all final state tracks and of the \Bs\ candidate, 
the angle between the \Bs\ momentum vector and the vector between PV and \Bs\ decay vertex, 
the vertex fit $\chisq$, the flight distance significance and transverse momentum of the \Bs\ candidate,  
and particle identification information of the muons and kaons in the final state. 

Several types of \bquark-hadron decays can mimic the final state of the signal decay and constitute potential sources of peaking background. 
The resonant decays $\BsToJPsiPhi$ 
and $\decay{\Bs}{\psitwos\phi}$ with $\decay{\psitwos}{\mup\mun}$ 
are rejected by applying vetoes on the dimuon mass regions around the charmonium resonances, $2946<m(\mup\mun)<3176\mevcc$ and $3592<m(\mup\mun)<3766\mevcc$. 
To account for the radiative tails of the charmonium resonances the vetoes are enlarged by $200\mevcc$ to lower $m(\mup\mun)$ for reconstructed \Bs\ masses below $5316\mevcc$.  
In the region $5416<m(\Bs)<5566\mevcc$ the vetoes are extended by $50\mevcc$ to higher $m(\mup\mun)$ to reject a small fraction of $\jpsi$ and $\psitwos$ decays that are misreconstructed at higher masses. 
The decay $\BdToKstmm$ ($\decay{\Kstarz}{\Kp\pim}$) can be reconstructed as signal if the pion is misidentified as a kaon. 
This background is strongly suppressed by particle identification criteria. 
In the narrow $\phi$ mass window, $2.4\pm 0.5$ misidentified $\BdToKstmm$ candidates are expected within $\pm 50\mevcc$ of the known \Bs\ mass of $5366\mevcc$~\cite{PDG2012}. 
The resonant decay $\BsToJPsiPhi$ can also constitute a source of peaking background if the $\Kp$~($\Km$) is misidentified as $\mup$~($\mun$) and vice versa. 
Similarly, the decay $\BdToJPsiKst$ ($\Kstarz\to \Kp\pi^-$) where the $\pi^-$~($\mun$) is misidentified as $\mun$~($\Km$) can mimic the signal decay. 
These backgrounds are rejected by requiring that the invariant mass of the $\Kp\mun$~($\Km\mup$) system, 
with kaons reconstructed under the muon mass hypothesis, 
is not within $\pm 50\mevcc$ around the known $\jpsi$ mass of $3096\mevcc$~\cite{PDG2012}, 
unless both the kaon and the muon pass stringent particle identification criteria. 
The expected number of background events from double misidentification in the \Bs\ signal mass region is $0.9 \pm 0.5$. 
All other backgrounds studied, including semileptonic 
$b\rightarrow c\,\mun\bar{\nu}_\mu(c\rightarrow s\,\mup\nu_\mu)$  
cascades, hadronic double misidentification from $\Bs\to \Dsm \pip (\Dsm\to\phi \pim )$, and the decay $\decay{\Lb}{\L(1520)\,\mup\mun}$, have been found to be negligible.

\section{Differential branching fraction}
\label{sec:differential}
\begin{figure}
\begin{center}
\includegraphics[width=12cm]{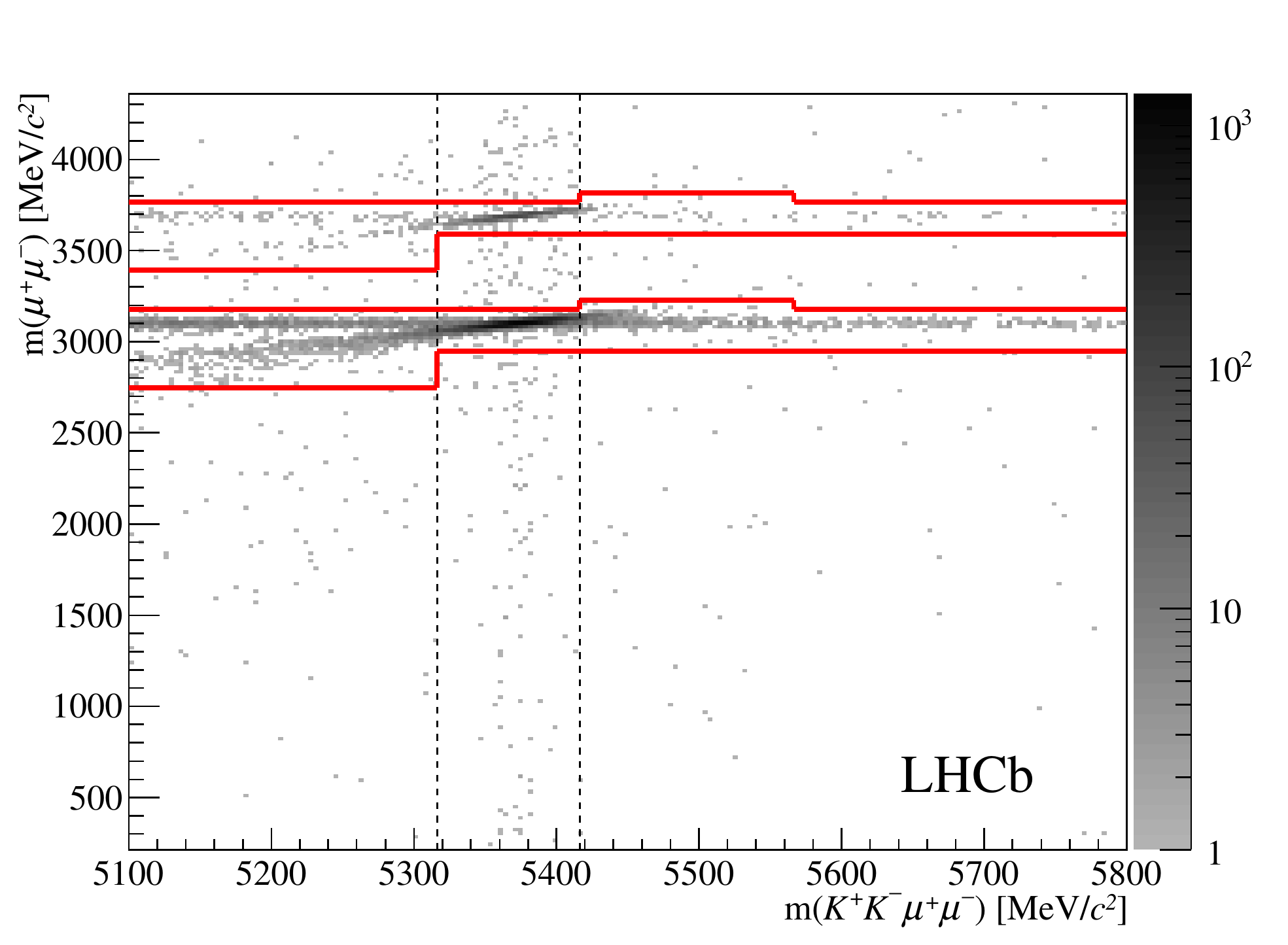}
\end{center}
\caption{\small 
Invariant $\mup\mun$ versus $\Kp\Km\mup\mun$ mass. 
The charmonium vetoes are indicated by the solid lines. 
The vertical dashed lines indicate the signal region of $\pm50\mevcc$ around the known \Bs\ mass 
in which the signal decay $\BsToPhimm$ is visible. 
}
\label{fig:fullselection}
\end{figure}

\begin{figure}
\begin{center}
\includegraphics[width=7.75cm]{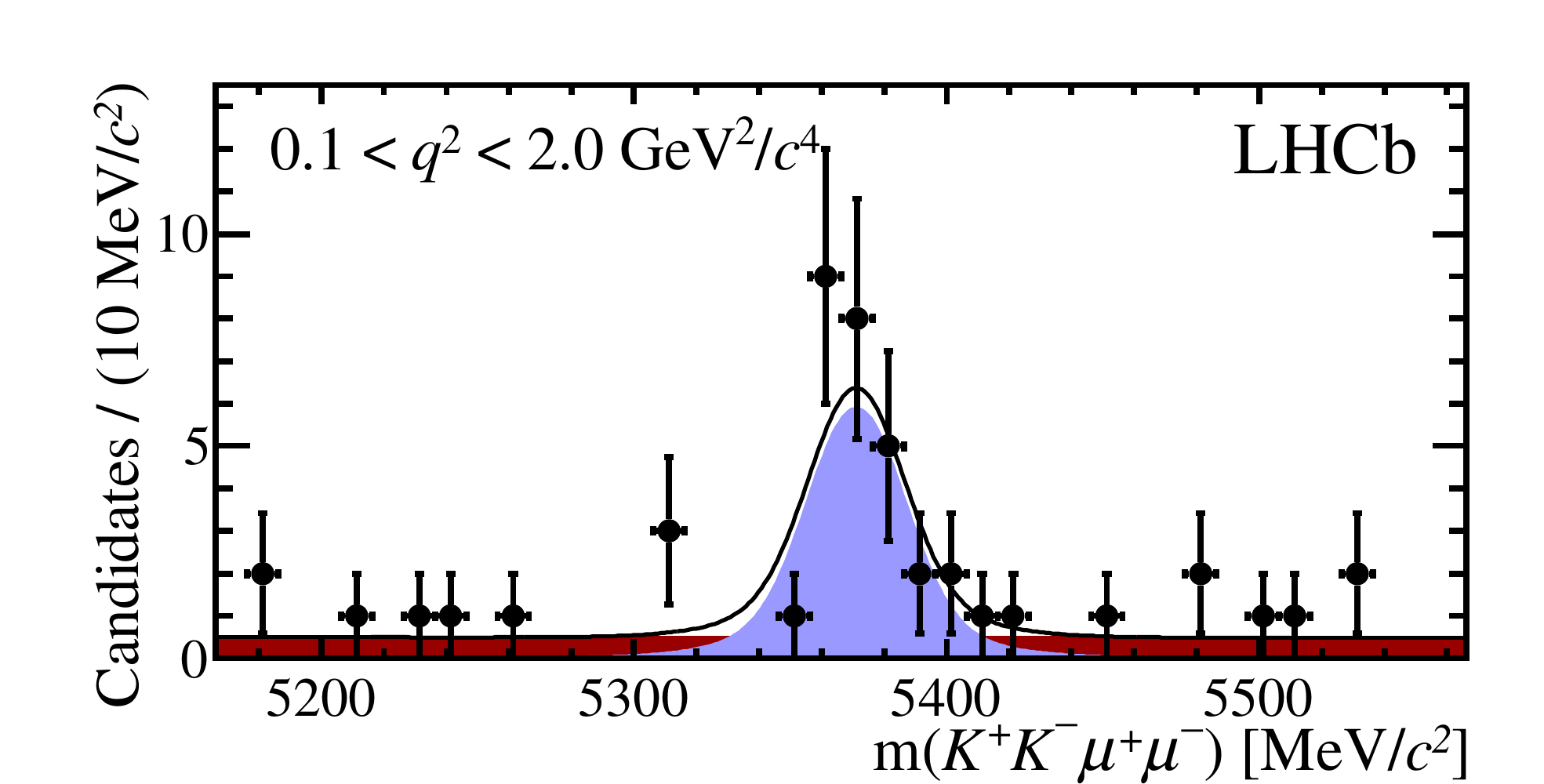}
\includegraphics[width=7.75cm]{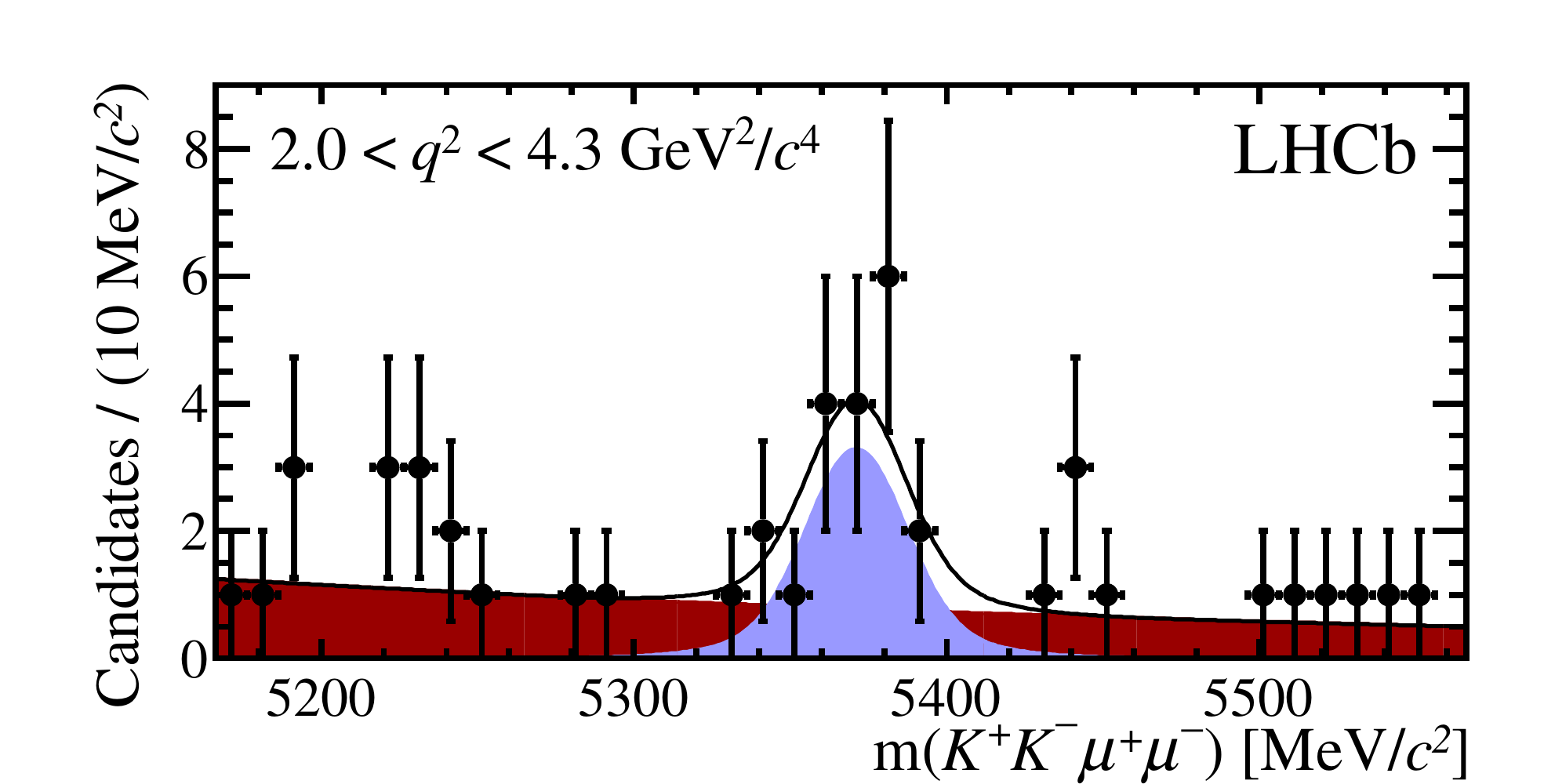}\\
\includegraphics[width=7.75cm]{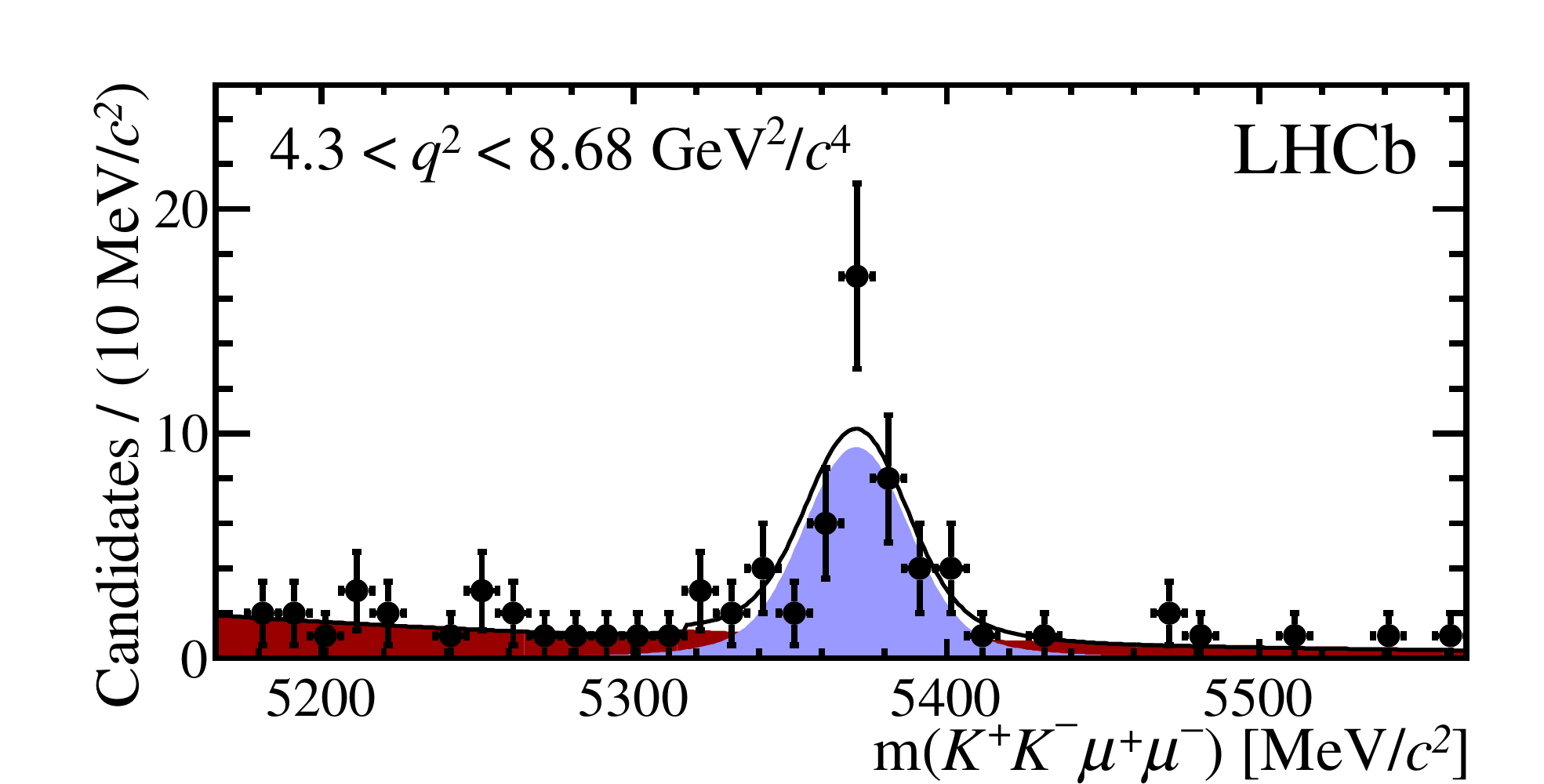}
\includegraphics[width=7.75cm]{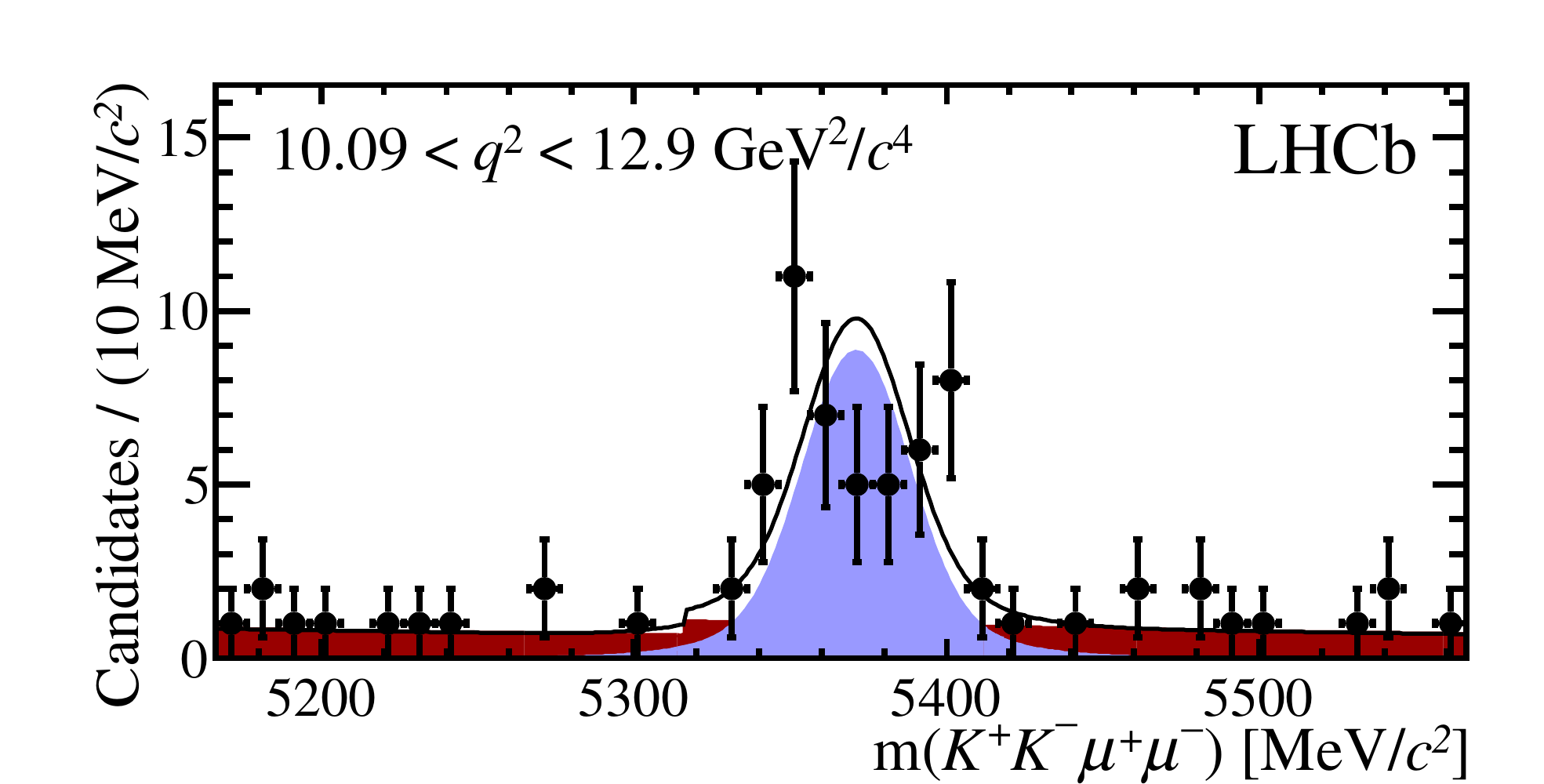}\\
\includegraphics[width=7.75cm]{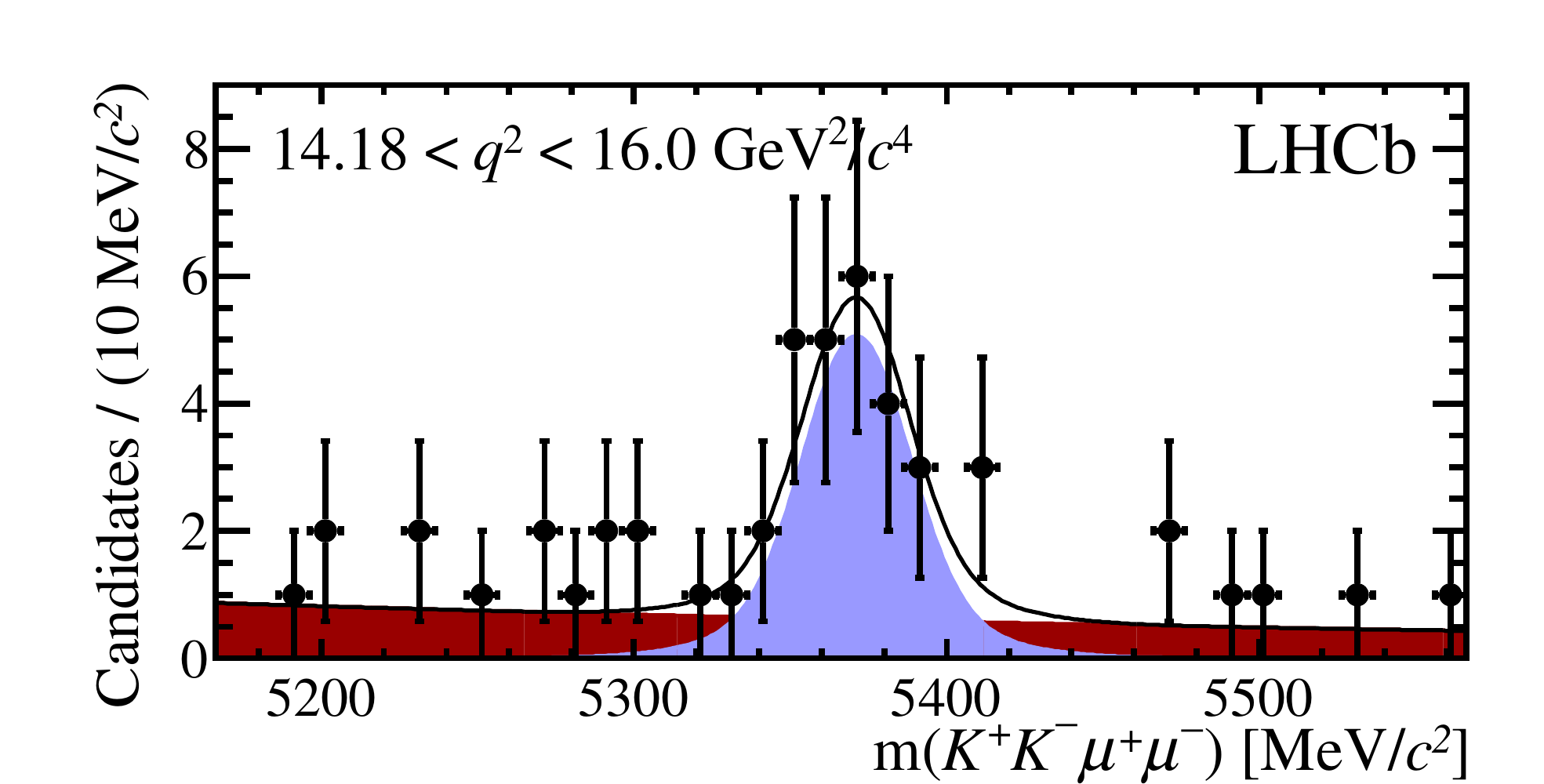}
\includegraphics[width=7.75cm]{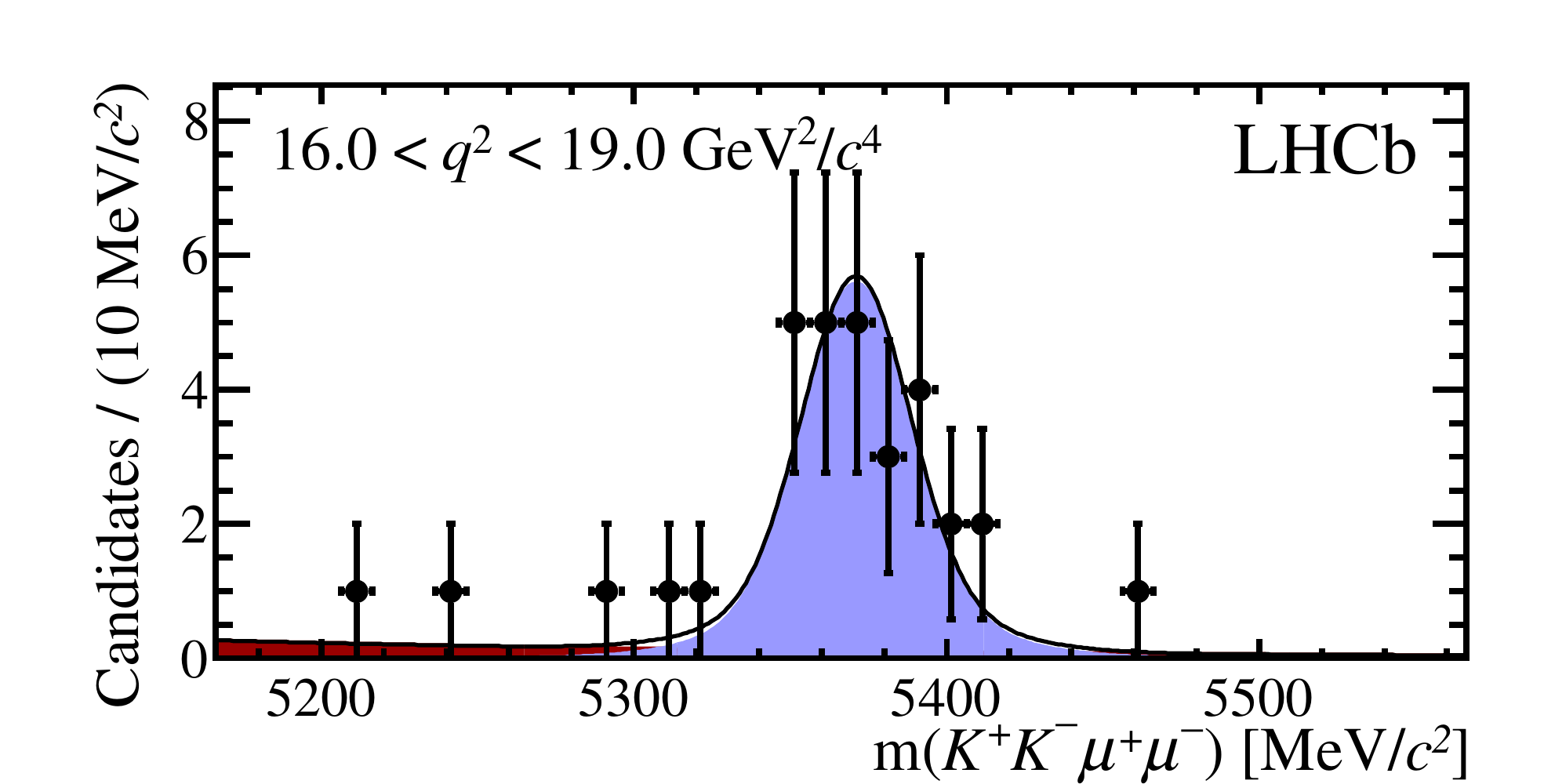}
\end{center}
\caption{\small
Invariant mass of $\Bs\to\phi\mup\mun$ candidates in six bins of invariant dimuon mass squared. 
The fitted signal component is denoted by the light blue shaded area, the combinatorial background component by the dark red shaded area. 
The solid line indicates the sum of the signal and background components. 
} 
\label{fig:diffnsig}
\end{figure}

Figure~\ref{fig:fullselection} shows the $\mup\mun$ versus the $\Kp\Km\mup\mun$ invariant mass of the selected candidates. 
The signal decay $\BsToPhimm$ is clearly visible in the \Bs\ signal region. 
The determination of the differential branching fraction is performed in six bins of $q^2$, given in Table~\ref{tab:diffresult}, and 
corresponds to the binning chosen for the analysis of the decay $\Bd\to\Kstarz\mup\mun$~\cite{bib:kstarmumu}. 
Figure~\ref{fig:diffnsig} shows the $\Kp\Km\mup\mun$ mass distribution in the six $q^2$ bins. 
The signal yields are determined by extended unbinned maximum likelihood fits to the reconstructed \Bs\ mass distributions. 
The signal component is modeled by a double Gaussian function. 
The resolution parameters are obtained from the resonant $\BsToJPsiPhi$ decay.  
A \mbox{$q^2$-dependent} scaling factor, determined with simulated $\BsToPhimm$ events, is introduced to account for the observed $q^2$ dependence of the mass resolution. 
The combinatorial background is described by a single exponential function. 
The veto of the radiative tails of the charmonium resonances is accounted for by using a scale factor. 
The resulting signal yields are given in Table~\ref{tab:diffresult}. 
Fitting for the signal yield over the full $q^2$ region, 
$174 \pm 15$ signal candidates are found. 
A fit of the normalisation mode $\Bs\to\jpsi\phi$ yields 
$(20.36\pm 0.14)\times 10^{3}$
candidates. 
\begin{table}
\caption{\small 
Signal yield and differential branching fraction \mbox{${\rm d}{\cal B}(\BsToPhimm)/{\rm d}q^2$} in six bins of $q^2$. 
Results are also quoted for the region $1<q^2<6\gevcc$ where theoretical predictions are most reliable. 
The first uncertainty is statistical, the second systematic, and the third from the branching fraction of the normalisation channel.}
\label{tab:diffresult}
\begin{center}
{\renewcommand*\arraystretch{1.25}
\begin{tabular}{r|c|c}
$q^2$ bin $(\gevgevcccc)$ & $N_{\rm sig}$ & ${\rm d}{\cal B}/{\rm d}q^2~(10^{-8}\gev^{-2}c^4)$\\ \hline
$0.10 < q^2 < \parbox{\widthof{12.90}}{2.00} $& $25.0\,^{+5.8}_{-5.2}$ & $ 4.72\,^{+1.09}_{-0.98}\pm 0.20 \pm 0.47 $ \\
$2.00 < q^2 < \parbox{\widthof{12.90}}{4.30}$& $14.3\,^{+4.9}_{-4.3}$ & $ 2.30\,^{+0.79}_{-0.69}\pm 0.11 \pm0.23 $ \\
$4.30 < q^2 < \parbox{\widthof{12.90}}{8.68}$& $41.2\,^{+7.5}_{-7.0}$ & $ 3.15\,^{+0.58}_{-0.53}\pm 0.12\pm 0.31 $ \\
$10.09 < q^2 < \parbox{\widthof{12.90}}{12.90}$& $40.7\,^{+7.7}_{-7.2}$ & $ 4.26\,^{+0.81}_{-0.75}\pm 0.26\pm 0.43 $ \\
$14.18 < q^2 < \parbox{\widthof{12.90}}{16.00}$& $23.8\,^{+5.9}_{-5.3}$ & $ 4.17\,^{+1.04}_{-0.93}\pm 0.24\pm 0.42 $ \\
$16.00 < q^2 < \parbox{\widthof{12.90}}{19.00}$& $26.6\,^{+5.7}_{-5.3}$ & $ 3.52\,^{+0.76}_{-0.70}\pm 0.20\pm 0.35 $ \\ \hline
$1.00 < q^2 < \parbox{\widthof{12.90}}{6.00}$& $31.4\,^{+7.0}_{-6.3}$ & $ 2.27\,^{+0.50}_{-0.46}\pm 0.11\pm 0.23 $ \\
\end{tabular}
}
\end{center}
\end{table}

\begin{figure}
\begin{center}
\includegraphics[width=10cm]{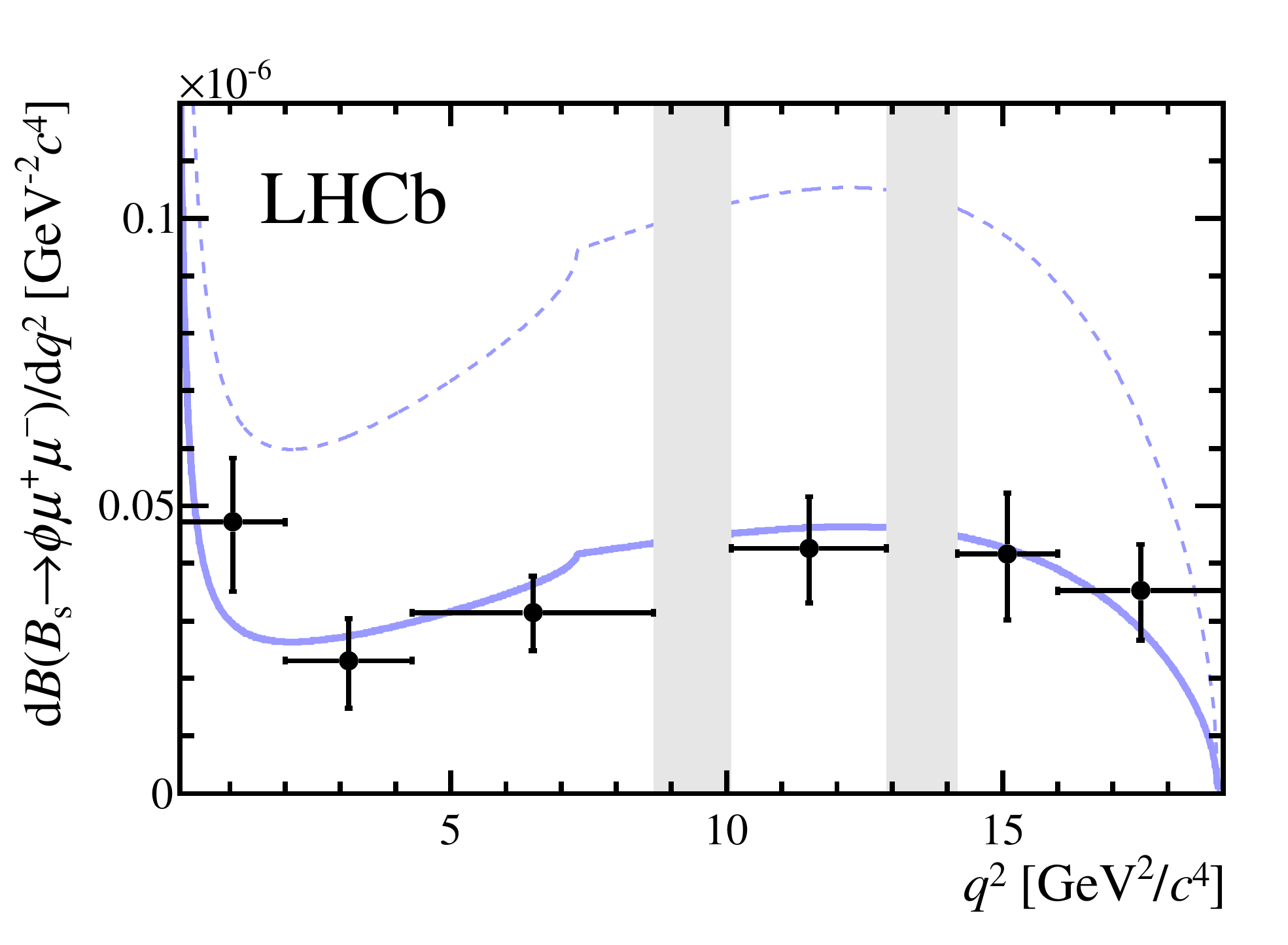}
\end{center}
\caption{\small 
Differential branching fraction ${\rm d}{\cal B}(\BsToPhimm)/{\rm d}q^2$. 
Error bars include both statistical and systematic uncertainties added in quadrature. 
Shaded areas indicate the vetoed regions containing the $\jpsi$ and $\psitwos$ resonances. 
The solid curve shows the leading order SM prediction, 
scaled to the fitted total branching fraction. 
The prediction uses the SM Wilson coefficients and leading order amplitudes given in Ref.~\cite{bib:altmannshofer}, 
as well as the form factor calculations in Ref.~\cite{bib:ball05}. 
\Bs\ mixing is included as described in Ref.~\cite{bib:hiller}. 
No error band is given for the theory prediction. 
The dashed curve denotes the leading order prediction scaled to a total branching fraction of $16\times 10^{-7}$~\cite{bib:smgeng}. 
}
\label{fig:diffbr}
\end{figure}

The differential branching fraction of the signal decay in the $q^2$ interval spanning from $q^2_{\rm min}$ to $q^2_{\rm max}$ 
is calculated according to
\begin{align}
\frac{{\rm d}{\cal B}(\BsToPhimm)}{{\rm d}q^2} &= \frac{1}{q^2_{\rm max}-q^2_{\rm min}}\frac{N_{\rm sig}}{N_{\jpsi\phi}}\frac{\epsilon_{\jpsi\phi}}{\epsilon_{\phi\mu^+\mu^-}} {\cal B}(\BsToJPsiPhi) {\cal B}(\decay{\jpsi}{\mup\mun}),~\label{eq:diffbr}
\end{align}
where $N_{\rm sig}$ and $ N_{\jpsi\phi}$ denote the yields of the $\BsToPhimm$ and $\BsToJPsiPhi$ candidates  
 and $\epsilon_{\phi\mu^+\mu^-}$ and $\epsilon_{\jpsi\phi}$ denote their respective efficiencies. 
Since the reconstruction and selection efficiency of the signal decay depends on $q^2$, a separate efficiency ratio $\epsilon_{\jpsi\phi}/\epsilon_{\phi\mu^+\mu^-}$ is determined for every $q^2$ bin. 
The branching fractions used in Eq.~\ref{eq:diffbr} are given by 
${\cal B}(\BsToJPsiPhi)=\left(10.50\pm 1.05\right)\times 10^{-4}$~\cite{bib:lhcbjpsiphi} 
and ${\cal B}(\decay{\jpsi}{\mup\mun})=\left(5.93\pm 0.06\right)\times 10^{-2}$~\cite{PDG2012}. 
The resulting $q^2$-dependent differential branching fraction ${\rm d}{\cal B}(\BsToPhimm)/{\rm d}q^2$ 
is shown in Fig.~\ref{fig:diffbr}. 
Possible contributions from \Bs\ decays to $\Kp\Km\mup\mun$, with the $\Kp\Km$ pair in an S-wave configuration, are neglected in this analysis. 
The S-wave fraction is expected to be small, for the decay $\Bs\to\jpsi \Kp\Km$ it is measured to be $(1.1\pm 0.1\,^{+0.2}_{-0.1})\%$~\cite{bib:lhcbjpsiphi} for the $\Kp\Km$ mass window used in this analysis. 

The total branching fraction is determined by summing the differential branching fractions in the six $q^2$ bins. 
Using the form factor calculations described in Ref.~\cite{bib:ball05} the signal fraction rejected by the charmonium vetoes is determined to be $17.7\%$. 
This number is confirmed by a different form factor calculation detailed in Ref.~\cite{bib:ali}. 
No uncertainty is assigned to the vetoed signal fraction. 
Correcting for the charmonium vetoes, the branching fraction ratio ${\cal B}\left(\BsToPhimm\right)/{\cal B}\left(\Bs\to\jpsi\phi\right)$ is measured to be
\begin{align*}
\frac{{\cal B}(\Bs\rightarrow\phi\mu^{+}\mu^{-})}{{\cal B}(\Bs\rightarrow\jpsi\phi)} &= \left(6.74\,^{+0.61}_{-0.56}\pm 0.16\right)\times 10^{-4}. 
\end{align*}
The systematic uncertainties will be discussed in detail in Sec.~\ref{sec:systdiffbr}. 
Using the known branching fraction of the normalisation channel the total branching fraction is 
\begin{align*}
{\cal B}(\BsToPhimm) &= \left(7.07\,^{+0.64}_{-0.59}\pm 0.17 \pm 0.71\right)\times 10^{-7},
\end{align*}
where the first uncertainty is statistical, the second systematic and the third from the uncertainty on the branching fraction of the normalisation channel. 

\subsection{Systematic uncertainties on the differential branching fraction}
\label{sec:systdiffbr}
The dominant source of systematic uncertainty 
on the differential branching fraction arises from the uncertainty on the branching fraction of the normalisation channel 
$\Bs\to\jpsi\phi$ ($\jpsi\to\mup\mun$), which is known to an accuracy of $10\%$~\cite{bib:lhcbjpsiphi}. 
This uncertainty is fully correlated between all $q^2$ bins. 

Many of the systematic uncertainties affect the relative efficiencies $\epsilon_{\jpsi\phi}/\epsilon_{\phi\mu^+\mu^-}$ that are determined using simulation. 
The limited size of the simulated samples 
causes an uncertainty of $\sim 1\%$ on the ratio in each bin. 
Simulated events are corrected for known discrepancies between simulation and data. 
The systematic uncertainties associated with these corrections 
(\eg\ tracking efficiency 
 and performance of the particle identification) 
are typically of the order of $1\text{--}2\%$. 
The correction procedure for the impact parameter resolution has an effect of up to $5\%$. 
Averaging the relative efficiency within the $q^2$ bins leads to a systematic uncertainty of $1\text{--}2\%$. 
Other systematic uncertainties of the same magnitude include the trigger efficiency and the uncertainties of the 
angular distributions of the signal decay $\Bs\rightarrow\phi\mu^+\mu^-$. 
The influence of the signal mass shape is found to be $0.5\%$. 
The background shape has an effect of up to $5\%$, which is evaluated by using a linear function to describe the mass distribution of the background instead of the nominal exponential shape. 
Peaking backgrounds cause a systematic uncertainty of $1\text{--}2\%$ on the differential branching fraction. 
The size of the systematics uncertainties on the differential branching fraction, added in quadrature, ranges from $4\text{--}6\%$. 
This is small compared to the dominant systematic uncertainty of $10\%$ due to the branching fraction of the normalisation channel, which is given separately in Table~\ref{tab:diffresult}, and the statistical uncertainty.

\section{Angular analysis}
\label{sec:angularanalysis}
The angular observables $F_{\rm L}$, $S_3$, $A_6$, and $A_9$ are determined using unbinned maximum likelihood fits to the distributions of 
$\ctk$, $\ctl$, $\Phi$, and the invariant mass of the $K^+K^-\mu^+\mu^-$ system. 
The detector acceptance and the reconstruction and selection of the signal decay distort the angular distributions 
given in Eqs.~\ref{eq:ctk}--\ref{eq:phi}. 
To account for this angular acceptance effect, an angle-dependent efficiency is introduced  
that factorises in $\ctk$ and $\ctl$,  
and is independent of the angle $\Phi$, \ie\ \mbox{$\epsilon(\ctk, \ctl, \Phi)=\epsilon_{K}(\ctk)\cdot\epsilon_{\ell}(\ctl)$}. 
The factors $\epsilon_{K}(\ctk)$ and $\epsilon_{\ell}(\ctl)$ are determined from fits to simulated events. 
Even Chebyshev polynomial functions of up to fourth order are used to parametrise $\epsilon_{K}(\ctk)$ and $\epsilon_{\ell}(\ctl)$ for each bin of $q^2$. 
The point-to-point dissimilarity method described in Ref.~\cite{bib:gof} confirms that the angular acceptance effect is well described by the acceptance model. 

Taking the acceptances into account and integrating Eq.~\ref{eq:untagged} over two angles, results in
\begin{align}
\frac{1}{{\rm d}\Gamma/{\rm d}q^2}\frac{{\rm d}^2\Gamma }{{\rm d}q^2\,{\rm d}\cos\theta_K} 
& = \epsilon_{K}(\ctk) \biggl[\frac{3}{4}(1-{F_{\rm L}}) (1 - \cos^2\theta_K)\,\xi_1 + \frac{3}{2} {F_{\rm L}} \cos^2\theta_K\,\xi_2\biggr],\label{eq:ctkacc}\\
\frac{1}{{\rm d}\Gamma/{\rm d}q^2}\frac{{\rm d}^2\Gamma }{{\rm d}q^2\,{\rm d}\cos\theta_\ell} 
& = \parbox{\widthof{$\epsilon_{K}(\ctk)$}}{$\epsilon_{\ell}(\ctl)$} \biggl[\frac{3}{8}(1-{F_{\rm L}}) ( 1+\cos^2\theta_\ell )\,\xi_3 + \frac{3}{4} {F_{\rm L}} ( 1-\cos^2\theta_\ell )\,\xi_4 \nonumber\\
& \hphantom{=\epsilon_{K}(\ctk)}  
+ \frac{3}{4} {A_6} \cos\theta_\ell\,\xi_3\biggr],\label{eq:ctlacc}\\
\frac{1}{{\rm d}\Gamma/{\rm d}q^2}\frac{{\rm d}^2\Gamma }{{\rm d}q^2\,{\rm d}\Phi} 
& = 
\biggl[\frac{1}{2\pi}\xi_1\xi_3 + \frac{1}{2\pi}F_{\rm L} (\xi_2\xi_4-\xi_1\xi_3) \nonumber\\
& 
\hphantom{={}}  
+\frac{1}{2\pi}{S_3}\cos2\Phi\,\xi_2\xi_3 + \frac{1}{2\pi} {A_9} \sin 2\Phi\,\xi_2\xi_3\biggr].\label{eq:phiacc} 
\end{align}
The terms $\xi_i$ are correction factors with respect to Eqs.~\ref{eq:ctk}--\ref{eq:phi} and are given by the angular integrals
\begin{align}
\xi_1 &= \frac{3}{8}\int_{-1}^{+1} (1+\cos^2\thetal )\epsilon_{\ell}(\ctl){\rm d}\ctl, \nonumber\\
\xi_2 &= \frac{3}{4}\int_{-1}^{+1} (1-\cos^2\thetal)\epsilon_{\ell}(\ctl){\rm d}\ctl, \nonumber\\
\xi_3 &= \frac{3}{4}\int_{-1}^{+1} (1-\cos^2\thetak )\epsilon_{K}(\ctk){\rm d}\ctk, \nonumber\\
\xi_4 &= \frac{3}{2}\int_{-1}^{+1} \cos^2\thetak \epsilon_{K}(\ctk){\rm d}\ctk 
.
\end{align}
Three two-dimensional maximum likelihood fits in the decay angles and the reconstructed \Bs\ mass are performed for each $q^2$ bin to determine the angular observables. 
The observable $F_{\rm L}$ is determined in the fit to the $\ctk$ distribution described by Eq.~\ref{eq:ctkacc}.  
The $\ctl$ distribution given by Eq.~\ref{eq:ctlacc} is used to determine $A_6$. 
Both $S_3$ and $A_9$ are measured from the $\Phi$ distribution, as described by Eq.~\ref{eq:phiacc}. 
In the fit of the $\Phi$ distribution a Gaussian constraint is applied to the parameter $F_{\rm L}$ using the value of $F_{\rm L}$ determined from the $\ctk$ distribution. 
The constraint on $F_{\rm L}$ has negligible influence on the values of $S_3$ and $A_9$. 
The angular distribution of the background events is fit using Chebyshev polynomial functions of second order. 
The mass shapes of the signal and background are described by the sum of two Gaussian distributions with a common mean, 
and an exponential function, respectively. 
The effect of the veto of the radiative tails on the combinatorial background is accounted for by using an appropriate scale factor. 

\begin{figure}
\begin{center}
\includegraphics[width=7.5cm]{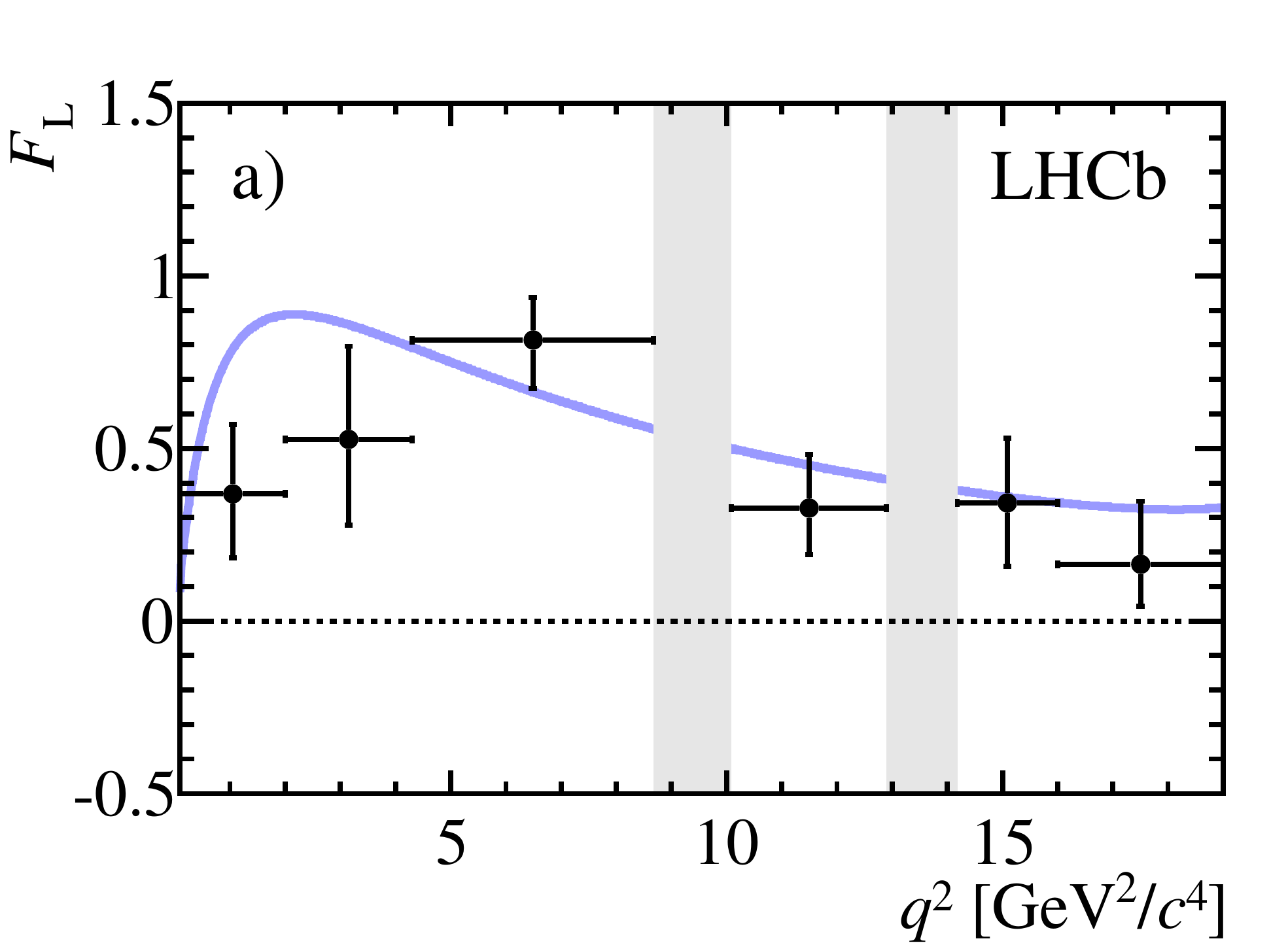}
\includegraphics[width=7.5cm]{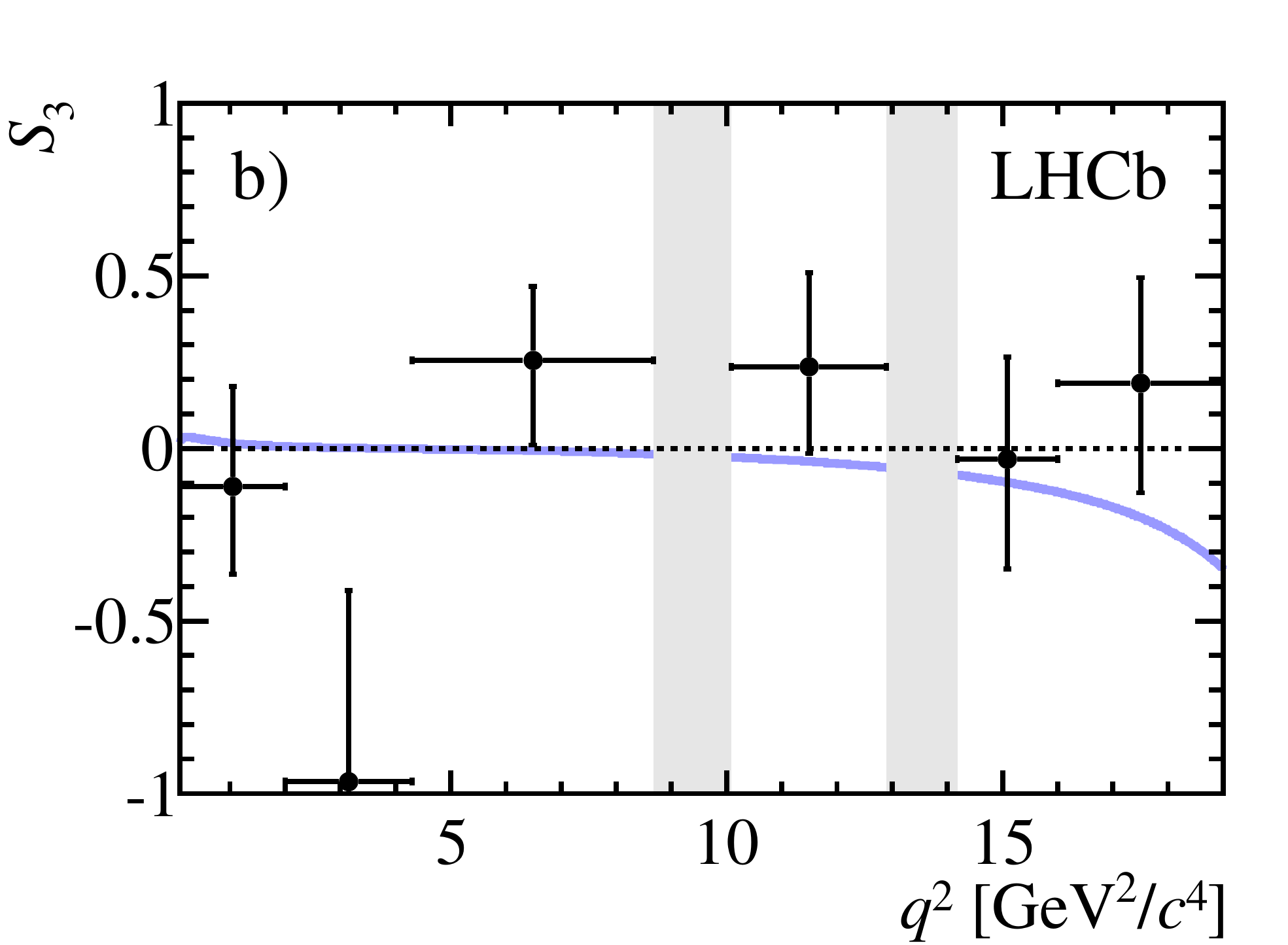}\\
\includegraphics[width=7.5cm]{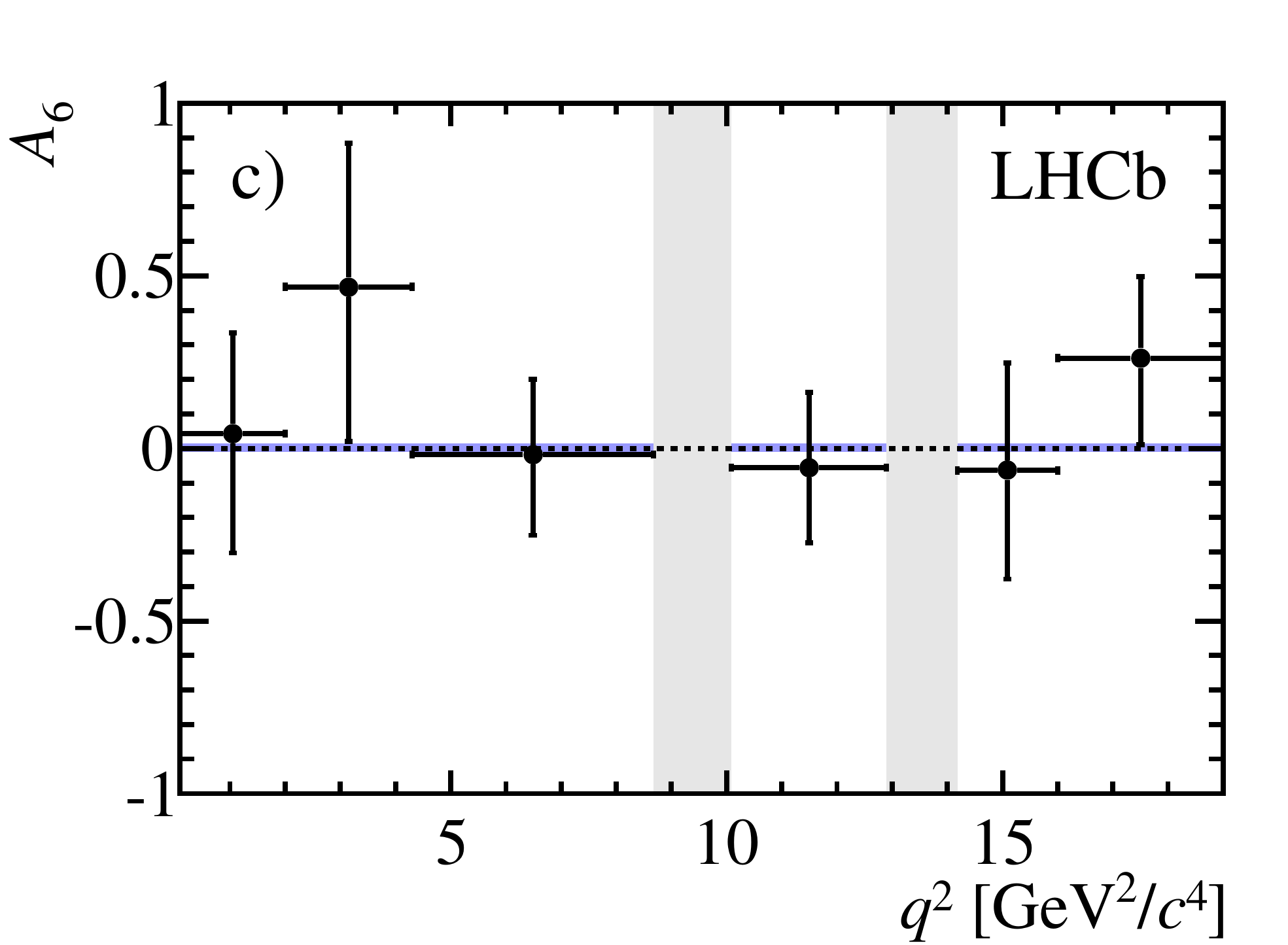}
\includegraphics[width=7.5cm]{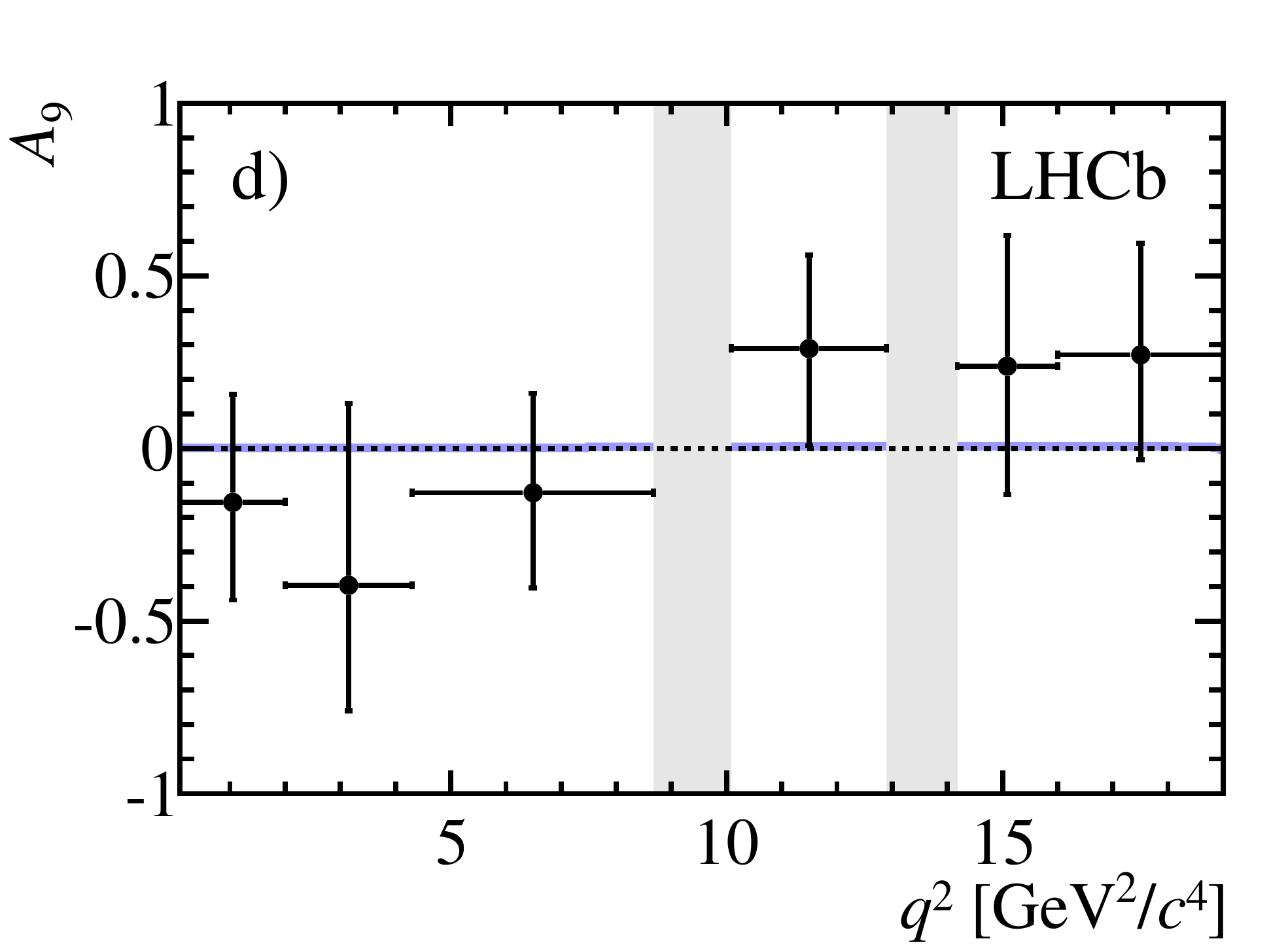}
\end{center}
\caption{\small 
a) Longitudinal polarisation fraction $F_{\rm L}$, b) $S_3$, c) $A_6$, and d) $A_9$ in six bins of $q^2$. 
Error bars include statistical and systematic uncertainties added in quadrature. 
The solid curves are the leading order SM predictions, 
using the Wilson coefficients and leading order amplitudes given in Ref.~\cite{bib:altmannshofer}, 
as well as the form factor calculations in Ref.~\cite{bib:ball05}. 
\Bs\ mixing is included as described in Ref.~\cite{bib:hiller}. 
No error band is given for the theory predictions. 
} 
\label{fig:angularresults}
\end{figure}

The measured angular observables are presented in Fig.~\ref{fig:angularresults} and Table~\ref{tab:angularresults}.  
The $68\%$ confidence intervals are determined using the Feldman-Cousins method~\cite{bib:fc}  
and the nuisance parameters are included using the plug-in method~\cite{bib:plugin}. 
\begin{table}
\caption{\small 
Results for the angular observables $F_{\rm L}$, $S_3$, $A_6$, and $A_9$ in bins of $q^2$. 
The first uncertainty is statistical, the second systematic.
}
\label{tab:angularresults}
\begin{center}
{\footnotesize\renewcommand*\arraystretch{1.25}
\begin{tabular}{r|r|r|r|r}
$q^2$ bin $(\gevgevcccc)$ & \multicolumn{1}{c|}{$F_{\rm L}$} & \multicolumn{1}{c|}{$S_3$} & \multicolumn{1}{c|}{$A_6$} & \multicolumn{1}{c}{$A_9$} \\ \hline
$0.10 < q^2 < \parbox{\widthof{12.90}}{2.00} $ & $0.37\,^{+0.19}_{-0.17} \pm 0.07$ & $-0.11\,^{+0.28}_{-0.25} \pm 0.05$ & $0.04\,^{+0.27}_{-0.32} \pm 0.12$ & $-0.16\,^{+0.30}_{-0.27} \pm 0.09$ \\
$2.00 < q^2 < \parbox{\widthof{12.90}}{4.30}$ & $0.53\,^{+0.25}_{-0.23} \pm 0.10$ & $-0.97\,^{+0.53}_{-0.03} \pm 0.17$ & $0.47\,^{+0.39}_{-0.42} \pm 0.14$ & $-0.40\,^{+0.52}_{-0.35} \pm 0.11$ \\
$4.30 < q^2 < \parbox{\widthof{12.90}}{8.68}$ & $0.81\,^{+0.11}_{-0.13} \pm 0.05$ & $0.25\,^{+0.21}_{-0.24} \pm 0.05$ & $-0.02\,^{+0.20}_{-0.21} \pm 0.10$ & $-0.13\,^{+0.27}_{-0.26} \pm 0.10$ \\
$10.09 < q^2 < \parbox{\widthof{12.90}}{12.90}$ & $0.33\,^{+0.14}_{-0.12} \pm 0.06$ & $0.24\,^{+0.27}_{-0.25} \pm 0.06$ & $-0.06\,^{+0.20}_{-0.20} \pm 0.08$ & $0.29\,^{+0.25}_{-0.26} \pm 0.10$ \\
$14.18 < q^2 < \parbox{\widthof{12.90}}{16.00}$ & $0.34\,^{+0.18}_{-0.17} \pm 0.07$ & $-0.03\,^{+0.29}_{-0.31} \pm 0.06$ & $-0.06\,^{+0.30}_{-0.30} \pm 0.08$ & $0.24\,^{+0.36}_{-0.35} \pm 0.12$ \\
$16.00 < q^2 < \parbox{\widthof{12.90}}{19.00}$ & $0.16\,^{+0.17}_{-0.10} \pm 0.07$ & $0.19\,^{+0.30}_{-0.31} \pm 0.05$ & $0.26\,^{+0.22}_{-0.24} \pm 0.08$ & $0.27\,^{+0.31}_{-0.28} \pm 0.11$ \\\hline
$1.00 < q^2 < \parbox{\widthof{12.90}}{6.00}$ & $0.56\,^{+0.17}_{-0.16} \pm 0.09$ & $-0.21\,^{+0.24}_{-0.22} \pm 0.08$ & $0.20\,^{+0.29}_{-0.27} \pm 0.07$ & $-0.30\,^{+0.30}_{-0.29} \pm 0.11$ \\
\end{tabular}
}
\end{center}
\end{table}

\subsection{Systematic uncertainties on the angular observables}
The dominant systematic uncertainty on the angular observables is due to the angular acceptance model. 
Using the point-to-point dissimilarity method detailed in Ref.~\cite{bib:gof}, the acceptance model is shown to describe the angular acceptance effect for simulated events 
at the level of $10\%$. 
A cross-check of the angular acceptance using the normalisation channel $\Bs\rightarrow\jpsi\phi$ shows good agreement of the angular observables with the values determined in Refs.~\cite{bib:cdfphis} and~\cite{bib:lhcbphis}. 
For the determination of the systematic uncertainty due to the angular acceptance model, variations of the acceptance curves are used that have the largest impact on the angular observables. 
The resulting systematic uncertainty is of the order of $0.05\text{--}0.10$, depending on the $q^2$ bin. 

The limited amount of simulated events accounts for a systematic uncertainty of up to $0.02$. 
The simulation correction procedure (for tracking efficiency, impact parameter resolution, and particle identification performance) has negligible effect on the angular observables. 
The description of the signal mass shape leads to a negligible systematic uncertainty. 
The background mass model causes an uncertainty of less than $0.02$. 
The model of the angular distribution of the background can have a large effect since the statistical precision of the background sample is limited. 
To estimate the effect, the parameters describing the background angular distribution are determined in the high \Bs\ mass sideband ($5416<m(\Kp\Km\mup\mun)<5566\mevcc$) 
using a relaxed requirement on the $\phi$ mass. 
The effect is typically $0.05\text{--}0.10$. 
Peaking backgrounds cause systematic deviations of the order of $0.01\text{--}0.02$. 
Due to the sizeable lifetime difference in the \Bs\ system~\cite{bib:lhcbphis}  
a decay time dependent acceptance can in principle affect the angular observables. 
The deviation of the observables due to this effect is studied and found to be negligible. 
The total systematic uncertainties, 
evaluated by adding all components in quadrature, 
are small compared to the statistical uncertainties.

\section{Conclusions}
\label{sec:conclusions}
The differential branching fraction of the FCNC decay $\Bs\rightarrow\phi\mu^{+}\mu^{-}$ has been determined. 
The results are summarised in Fig.~\ref{fig:diffbr} and in Table~\ref{tab:diffresult}. 
Using the form factor calculations in Ref.~\cite{bib:ball05} 
to determine the fraction of events removed by the charmonium vetoes, 
the relative branching fraction ${\cal B}(\Bs\rightarrow\phi\mu^{+}\mu^{-})/{\cal B}(\Bs\rightarrow\jpsi\phi)$ is determined to be
\begin{eqnarray*}
\frac{{\cal B}(\Bs\rightarrow\phi\mu^{+}\mu^{-})}{{\cal B}(\Bs\rightarrow\jpsi\phi)} &=& \left(6.74\,^{+0.61}_{-0.56}\pm 0.16\right)\times 10^{-4}. 
\end{eqnarray*}
This value is 
compatible with 
a previous measurement by the CDF collaboration of \mbox{${\cal B}(\Bs\rightarrow\phi\mu^{+}\mu^{-})/{\cal B}(\Bs\rightarrow\jpsi\phi)= \left(11.3 \pm 1.9\pm 0.7\right)\times 10^{-4}$}~\cite{bib:cdfphimumu}  
and a recent preliminary result which yields 
${\cal B}(\Bs\rightarrow\phi\mu^{+}\mu^{-})/{\cal B}(\Bs\rightarrow\jpsi\phi)=\left(9.0 \pm 1.4 \pm 0.7\right)\times 10^{-4}$~\cite{bib:cdfphimumunew}. 
Using the branching fraction of the normalisation channel, 
${\cal B}(\BsToJPsiPhi)=\left(10.50\pm 1.05\right)\times 10^{-4}$~\cite{bib:lhcbjpsiphi},  
the total branching fraction of the decay is determined to be
\begin{eqnarray*}
{\cal B}(\Bs\rightarrow\phi\mu^{+}\mu^{-}) &=& \left(7.07\,^{+0.64}_{-0.59} \pm 0.17 \pm 0.71\right)\times 10^{-7}, 
\end{eqnarray*}
where the first uncertainty is statistical, the second systematic, and the third from the uncertainty of the branching fraction of the normalisation channel. 
This measurement constitutes the most precise determination of the $\Bs\rightarrow\phi\mu^{+}\mu^{-}$ branching fraction to date. 
The measured value is lower than the SM theory predictions that range from $14.5\times 10^{-7}$ to $19.2\times 10^{-7}$~\cite{bib:smgeng, bib:smerkol, bib:smyilmaz, bib:smchang}. 
The uncertainties on these predictions originating from the form factor calculations are typically of the order of $20\text{--}30\%$. 

In addition, the first angular analysis of the decay $\Bs\rightarrow\phi\mu^{+}\mu^{-}$ has been performed. 
The angular observables $F_{\rm L}$, $S_3$, $A_6$, and $A_9$ are determined in bins of $q^2$, using the distributions of 
$\ctk$, $\ctl$, and $\Phi$. 
The results are summarised in Fig.~\ref{fig:angularresults}, and the numerical values are given in Table~\ref{tab:angularresults}. 
All measured angular observables are consistent with the leading order SM expectation.

\section*{Acknowledgements}

\noindent We express our gratitude to our colleagues in the CERN
accelerator departments for the excellent performance of the LHC. We
thank the technical and administrative staff at the LHCb
institutes. We acknowledge support from CERN and from the national
agencies: CAPES, CNPq, FAPERJ and FINEP (Brazil); NSFC (China);
CNRS/IN2P3 and Region Auvergne (France); BMBF, DFG, HGF and MPG
(Germany); SFI (Ireland); INFN (Italy); FOM and NWO (The Netherlands);
SCSR (Poland); ANCS/IFA (Romania); MinES, Rosatom, RFBR and NRC
``Kurchatov Institute'' (Russia); MinECo, XuntaGal and GENCAT (Spain);
SNSF and SER (Switzerland); NAS Ukraine (Ukraine); STFC (United
Kingdom); NSF (USA). We also acknowledge the support received from the
ERC under FP7. The Tier1 computing centres are supported by IN2P3
(France), KIT and BMBF (Germany), INFN (Italy), NWO and SURF (The
Netherlands), PIC (Spain), GridPP (United Kingdom). We are thankful
for the computing resources put at our disposal by Yandex LLC
(Russia), as well as to the communities behind the multiple open
source software packages that we depend on.

\addcontentsline{toc}{section}{References}
\bibliographystyle{LHCb}
\bibliography{main,LHCb-PAPER,LHCb-CONF}

\end{document}